\documentclass[twocolumn,aps,floatfix,superscriptaddress,longbibliography]{revtex4-1}
\pdfoutput=1 

\usepackage{color}
\usepackage{graphicx}
\usepackage{subfigure}
\usepackage{afterpage}
\usepackage{dcolumn}
\usepackage{bm}
\usepackage{color}
\usepackage{float}
\usepackage{amsmath}
\usepackage{bm}
\usepackage{amssymb}
\usepackage{siunitx}
\usepackage{natbib}
\usepackage{tikz}
\usepackage{hhline}

\usepackage[colorlinks=true, urlcolor=blue, anchorcolor=blue, citecolor=blue,filecolor=blue,linkcolor=blue,menucolor=blue]{hyperref}

\def\url#1{\expandafter\string\csname #1\endcsname}

\def\G{G}

\newcounter{defcounter}
\begin{document}

\title{Mixed phases in feedback Ising models}

\author{Yi-Ping Ma}
\email[Contact author: ]{yiping.ma@northumbria.ac.uk}
\affiliation{Department of Mathematics, Physics and Electrical Engineering, Northumbria University, Newcastle upon Tyne, NE1 8ST, UK}

\author{Ivan Sudakow}
\email [Contact author: ]{ivan.sudakow@open.ac.uk}
\affiliation{School of Mathematics and Statistics, The Open University, Milton Keynes, MK7 6AA, UK}

\author{P. L. Krapivsky}
\email[Contact author: ]{pkrapivsky@gmail.com}
\affiliation{Department of Physics, Boston University, Boston, Massachusetts 02215, USA}
\affiliation{Santa Fe Institute, Santa Fe, New Mexico 87501, USA}

\begin{abstract}
We study mean-field Ising models  in which the coupling depends on the magnetization via a feedback function. We identify mixed phases (MPs) and show that they can be stable at zero temperature for sufficiently strong feedback. Moreover, stable MPs are always super-stable, meaning that perturbations decay linearly in time. Feedback Ising models (FIMs) provide a useful framework for phase transformations between aligned phases via stable and unstable intermediate phases in multistable systems. We also analyze the dynamical behavior of FIMs driven by a varying magnetic field and discuss basic properties of finite-dimensional FIMs.
\end{abstract}

\maketitle

The concepts of ordered phases and phase transitions originated in the context of ferromagnetism \cite{Sethna}, and the Ising model (IM) played a central role in these developments. The IM also provides an excellent framework for investigating dynamics \cite{Glauber63,Bray94,krap2010}. The mean-field version of the IM, the Curie-Weiss model (CWM), is tractable and captures the chief behavior: Alignment of the spins caused by the ferromagnetic interaction prevails over thermal noise below a critical temperature, and results in an ordered state in which the average magnetization is non-zero. The zero-temperature dynamics of the IM are understood in the mean-field limit and in two dimensions \cite{2D-IM-Kip,2D-IM-Jason} due to the intriguing connection with critical percolation, but remain largely unexplained in the physically interesting three-dimensional case \cite{IM,IM-3d,IM-super}.

The IM was extended beyond idealized magnetic materials and applied to subjects ranging from neuroscience and psychology \cite{Roud09,Cramer16} to atmospheric science and finance \cite{Majda2012, econ}. Feedback is an indispensable feature in most of these subjects, yet it is absent in the CWM. In neural tissue, Hebbian plasticity makes the synaptic weight between two neurons proportional to their co-activity; the resulting Hopfield network is an Ising system with state-dependent couplings that stabilize multiple memory attractors \cite{Hopfield1982}. On a planetary scale, the ice-albedo feedback couples radiative balance to sea ice topography \cite{MaSudakov2019}.  Ecological regime shifts \cite{Noble2018} and collective decision-making \cite{Grabowski2006} display analogous loops that either reinforce or oppose the prevailing order. Our goal is to enrich the CWM with feedback. The equilibrium and dynamical characteristics of the feedback Ising model (FIM) are nontrivial already at zero temperature; in this paper, we focus on this extreme case.

The IM can be defined on any graph: There is a spin $s_i=\pm 1$ at each vertex and neighboring spins interact, contributing $J s_i s_j$ to the energy. The classical IM imposes a uniform external magnetic field $h$ on each spin and a uniform ferromagnetic coupling $J$ set to unity without loss of generality. The inverse temperature is $\beta>0$, and the order parameter is the magnetization $m=\langle s_i\rangle$. Feedback arises when a control parameter becomes a function of the order parameter, thereby extending the IM beyond 2-spin interactions.

Replacing 2-spin interactions by $p$-spin interactions is popular in spin glasses \cite{mezard2009information}. In combinatorial optimization problems, the dire consequences of first-order quantum phase transitions for quantum annealing \cite{jorg2010energy} can be circumvented by reverse annealing in the fully connected $p$-spin model \cite{ohkuwa2018reverse,yamashiro2019dynamics}. In machine learning, $p$-neuron interactions enable large capacity for dense associative memory \cite{krotov2016dense}.  The $p$-spin models typically consider a single $p$ rather than multiple $p$'s. 

There were a few attempts to incorporate feedback into the IM, e.g., in modelling markets \cite{Bornholdt2001,krause2012opinion}. In a different context, linear feedback inserted into mean-field Landau theory converts a static critical point into an Andronov-Hopf bifurcation, yielding macroscopic limit-cycle oscillations confirmed for a fully connected Ising system \cite{DeMartino2019}. These studies incorporate feedback into the external field or the temperature, but not the coupling.

In contrast, we assume that the feedback causes the coupling between two neighboring spins to depend on the magnetization $m$. The resulting FIM is defined by the Hamiltonian
\begin{equation}
\label{ham-F}
H_\text{FIM}=-h\sum_is_i- \frac{1}{N}f(m)\sum_{\langle i,j\rangle} s_i s_j,
\end{equation}
where $\langle i,j\rangle$ indicates that sites $i$ and $j$ are neighbors. 
The microscopic origin of $f(m)$ characterizing feedback  --- synaptic plasticity, radiative transfer, resource limitation --- is system-specific, but its theoretical consequences are universal. We remark that the mean-field FIM with $f(m)$  linear in magnetization can be interpreted as the IM with 2-spin and 3-spin interactions:
\begin{equation}
\label{Ham:23}
H_{23}=-h\sum_is_i-\frac{J_2}{N}\sum_{i<j} s_i s_j -\frac{J_3}{N^2}\sum_{i<j<k} s_i s_j s_k.
\end{equation}
Equilibrium characteristics of the general FIM, Eq.~(\ref{ham-F}), are analytically tractable: one can derive bifurcation diagrams, Maxwell constructions, and dynamic trapping criteria in closed form, providing a benchmark against which more elaborate feedback systems (e.g., neural, climatic, or socio-economic) can be calibrated.

We will explore the dynamical properties of the FIM, but we begin by mentioning earlier work on equilibrium characteristics. If $f(m)$ is expanded as a Taylor series in $m$, the Hamiltonian (\ref{ham-F}) becomes a weighted sum of $p$-spin interactions, also known as equivalent-neighbor or separable interactions in the early literature. The free energy of systems with separable interactions \cite{Capel1976}, including separable short-ranged (SR) interactions \cite{Ouden1976}, can be derived using a minimax principle \cite{Bogolyubov1972}. A convex-envelope construction \cite{Capel1977} has been devised to reveal the critical behavior of systems with separable SR interactions under small perturbations \cite{Capel1979}. 

The mixed $p$-spin models (MpMs) typically include either 3-spin or 4-spin interactions in addition to the classical IM. In finite dimensions, 4-spin interactions arise in an exactly solved MpM with a nearest-neighbor pair interaction and an infinite-range pair-pair interaction \cite{oitmaa1975critical}; the exactly solved Baxter-Wu model uses 3-spin interactions on the triangular lattice \cite{Baxter_Wu}. In the mean-field limit, the equilibrium properties are analyzed for the MpMs with 4-spin interactions \cite{thompson1974model,Capel1976}, possibly with $2n$-spin interactions for $n\geq3$ \cite{mckerrell1972critical, bowers1972first}, while the MpM with 3-spin interactions can serve as a mean-field approximation to the Baxter-Wu model with additional 2-spin interactions \cite{Hemmer}. Recently, this 3-spin MpM has been applied to human-AI ecosystems \cite{AI-Human}, prompting further analysis of its equilibrium properties \cite{CW-23,CW-3}. Our FIM is the most general MpM in the mean-field limit. Besides, it generates a rich class of MpMs in finite dimensions.

{\it Feedback Ising model.}---We focus on the FIM in the $N\to\infty$ limit. The rescaled Hamiltonian reads
\begin{equation}
\label{eq:hat-H}
    {\cal H}(m)=\lim_{N\to\infty}\frac{1}{N}H_\text{FIM}=-hm-\frac{1}{2}f(m)m^2.
\end{equation}

The simplest feedback coupling is $f(m)=1+\gamma m$, where $\gamma\geq 0$ measures the feedback force. The negative feedback, $\gamma<0$, is equivalent to the positive feedback by flipping the signs of $s_i$, $h$, and $\gamma$ simultaneously in the Hamiltonian in Eq.~(\ref{ham-F}). When $\gamma=0$, we recover the classical IM Hamiltonian. When $\gamma=1$, the coupling $f(m)$ becomes twice the fraction of up-spins, which can be desirable for modeling purposes. We tacitly assume that $0\leq \gamma\leq 1$, so that the coupling $f(m)$ is overall ferromagnetic. We occasionally comment on the systems with $\gamma>1$ in which the coupling $f(m)$ is antiferromagnetic for $m\in[-1,-1/\gamma)$. An alternative cutoff model where $f(m)=0$ for $m\in[-1,-1/\gamma)$ is left for future work.

Our strategy is to first analyze FIMs with a general feedback function $f$ (general FIMs) and then apply this general theory to the FIM with a linear $f$ (linear FIM). The magnetic field $h$ is a control parameter, which can be time-dependent. We focus on the zero-temperature limit where spin flips are deterministic. This limit is relevant in many real-world applications where the temperature is hard to define or estimate.

The key feature of the FIM at zero temperature is that stable mixed phases (MPs) with $|m|<1$ can exist even if the coupling is ferromagnetic, i.e., $f(m)>0$ for all $|m|\leq1$. In the classical IM, stable MPs are common at finite temperatures due to thermal fluctuations; at zero temperature, they only exist for antiferromagnetic coupling $f(m)=-1$, not for ferromagnetic coupling $f(m)=1$. We also show that stable MPs are always super-stable, i.e., any perturbation decays linearly and vanishes in finite time. In a time-dependent magnetic field $h(t)$, this super-stability implies that the system is trapped at stable MP unless $|h'(t)|$ is large enough.

We identify four types of phase transitions at branch endpoints on the bifurcation diagram and find that two of them are first-order while the other two are second-order as $h'(t)\to0$. Within each pair, one type is second-order while the other type is third-order for $|h'(t)|$ small. As $h$ varies globally between $\pm\infty$, the linear FIM exhibits a three-stage phase transition, while general FIMs can exhibit an arbitrary number of intermediate stages.

{\it Glauber dynamics.}---Under Glauber dynamics \cite{Glauber63,Bray94,krap2010}, the time evolution of the magnetization $m$ is given by 
\begin{equation}
\label{eq:dmdt-general}
    \frac{dm}{dt}=-m+\tanh(-\beta {\cal H}'(m))\equiv a(m),
\end{equation}
where $\beta$ is the inverse temperature, and the derivative ${\cal H}'=\frac{d {\cal H}}{dm}$ of the rescaled Hamiltonian 
\eqref{eq:hat-H} is 
\begin{subequations}
\begin{align}
\label{eq:hat-H-prime}
    & {\cal H}'(m)=-h+g(m), \\
    &g(m)\equiv-mf(m)-\frac{1}{2}m^2f'(m).
\end{align}
\end{subequations}
The logarithm of the probability density function $\rho(m)$ at thermodynamic equilibrium is given by
\begin{subequations}
\begin{align}
\label{eq:Um}
   & -\frac{\log(\rho(m))}{N}\to\beta{\cal H}(m)+V(m)\equiv U(m), \\
    & V(m)\equiv m\cdot \text{arctanh}(m) + \frac{1}{2} \log(1 - m^2).
\end{align}
\end{subequations}
The potential $U(m)$ is also known as the information content or surprisal in information theory. Thermodynamically, $U(m)/\beta$ is known as the free energy. The global minimum of $U(m)$ represents the ground state of the system; local minima of $U(m)$ are metastable states.

In the zero-temperature limit  ($\beta\to+\infty$), the function $\tanh(\beta x)$ tends to $\textrm{sgn}(x)$ whose discontinuity at $x=0$ is a primary source of zero-temperature phase transitions. Also, the potential $U(m)$ tends to $\beta {\cal H}(m)$, so transitions between stable equilibria are exponentially slow.

{\it Equilibria and their stability.}---In the $\beta\to+\infty$ limit, Eq.~(\ref{eq:dmdt-general}) becomes
\begin{equation}
\label{eq:dmdt-general-0}
    \frac{dm}{dt}=-m+\text{sgn}(h(t)-g(m)).
\end{equation}
In the stripe ${\cal S}=\{(h,m):|m|\leq1\}$ on the $(h,m)$-plane, there are an upper branch of stable equilibria
\begin{equation}
    {\cal C}_+=\{(h,m):m=m_+\equiv+1,h>h_+\equiv g(1)\},
\end{equation}
a lower branch of stable equilibria
\begin{equation}
    {\cal C}_-=\{(h,m):m=m_-\equiv-1,h<h_-\equiv g(-1)\},
\end{equation}
and a middle curve of equilibria
\begin{equation}
    {\cal C}_0=\{(h,m):|m|<1,h=g(m)\}.
\end{equation}
The ${\cal C}_0$ curve contains MPs, i.e., equilibria with $|m|<1$, which can be either stable or unstable. Depending on $g(m)$, ${\cal C}_0$ can consist of branch(es) separated by fold(s).

If we divide the stripe ${\cal S}$ into two regions
\begin{equation*}
    {\cal R}_-=\{(h,m):h<g(m)\},\quad{\cal R}_+=\{(h,m):h>g(m)\}
\end{equation*}
separated by the curve ${\cal C}_0$, then Eq.~(\ref{eq:dmdt-general-0}) simplifies to
\begin{equation}\label{eq:dmdt-gen-sim}
    \frac{dm}{dt}=\begin{cases}
        -m-1, & (h(t),m)\in{\cal R}_-\\
        -m+1, & (h(t),m)\in{\cal R}_+
    \end{cases},
\end{equation}
whose general solution is
\begin{equation}
    m(t)=\begin{cases}
        -1+c_-e^{-t}, & (h(t),m)\in{\cal R}_-\\
        1+c_+e^{-t}, & (h(t),m)\in{\cal R}_+
    \end{cases},
\end{equation}
where $c_\pm$ are constants. Thus, any initial condition on ${\cal C}_\pm$ remains on ${\cal C}_\pm$ since the equilibria $m_\pm$ are exponentially stable. However, an initial condition $m(t_0)=m_0$ on ${\cal C}_0$, i.e., with $h(t_0)=g(m_0)$, can either enter ${\cal R}_{\text{sgn}(h'(t_0))}$ or remain on ${\cal C}_0$. When $h'(t_0)>0$, ${\cal R}_+$ is entered when
\begin{equation}
    h'(t_0)>m'(t_0)g'(m_0)=(-m_0+1)g'(m_0).
\end{equation}
When $h'(t_0)<0$, ${\cal R}_-$ is entered when
\begin{equation}
    h'(t_0)<m'(t_0)g'(m_0)=(-m_0-1)g'(m_0).
\end{equation}
Thus, $m_0$ is super-unstable when $g'(m_0)<0$ for any $h'(t_0)$ since perturbation grows linearly in $t$ with rates
\begin{equation}
    a(m_0^-)=-m_0-1,\quad a(m_0^+)=-m_0+1.
\end{equation}
On the other hand, $m_0$ is super-stable when $g'(m_0)>0$ and $h'(t_0)$ satisfies a trapping condition
\begin{equation}\label{eq:trap-cond}
    h'(t_0)\in((-m_0-1)g'(m_0),(-m_0+1)g'(m_0)),
\end{equation}
since perturbation decays linearly in $t$ with rates
\begin{equation}
    a(m_0^-)=-m_0+1,\quad a(m_0^+)=-m_0-1.
\end{equation}
If Eq.~(\ref{eq:trap-cond}) is violated, then the rapid rise or fall in the magnetic field throws the trajectory off equilibrium.

\begin{figure}
    \centering
    \subfigure[]{\includegraphics[width=.21\textwidth]{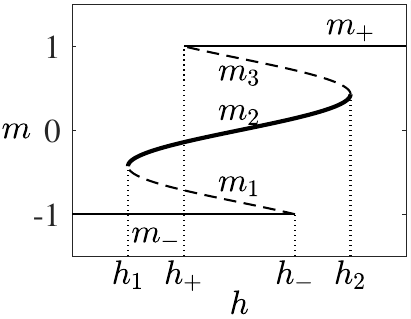}\label{fig:FIM_illu_s1}}
    \subfigure[]{\includegraphics[width=.21\textwidth]{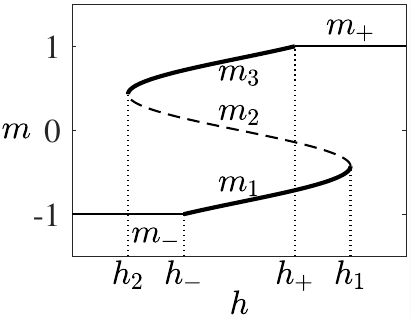}\label{fig:FIM_illu_s-1}}
    \caption{An illustration of possible bifurcation diagrams of equilibria on the $(h,m)$-plane in general FIMs with (a) $g'(m_-)<0$; and (b) $g'(m_-)>0$. Solid branches are stable, bold branches are super-stable, and dashed branches are unstable. Here, $g$ can be any continuous function; see End Matter for further discussions and choices of $g$ in each panel.}
    \label{fig:FIM_illu}
\end{figure}

For a static magnetic field, i.e., $h'(t_0)=0$, an MP $m_0$ is super-stable when $g'(m_0)>0$ and super-unstable when $g'(m_0)<0$, i.e., the stability of the ${\cal C}_0$ curve changes at every fold. Thus, if we denote $h$ at the $i$-th fold as $h_i$ and the branch below the $i$-th fold as $m_i$, $i=1,2,\cdots$, counting $(h_+,m_+)$ as the last fold, then the $m_i$ branch is super-stable for even $i$ when $g'(m_-)<0$ and for odd $i$ when $g'(m_-)>0$; see Fig.~\ref{fig:FIM_illu} for an illustration.

\begin{figure}
    \centering
    \subfigure[]{\includegraphics[width=.21\textwidth]{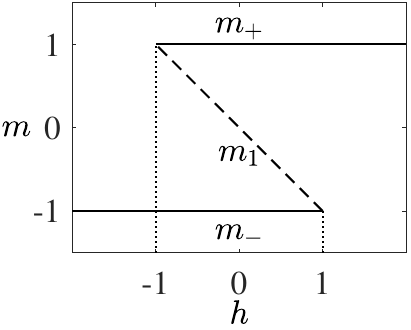}\label{fig:IM_bif}}
    \subfigure[]{\includegraphics[width=.21\textwidth]{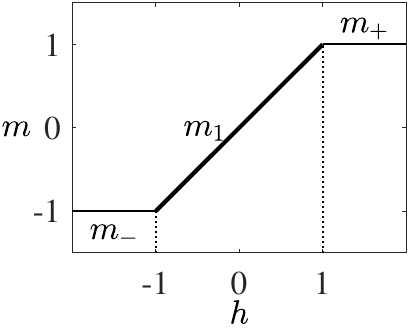}\label{fig:IM_bif_anti}}
    \subfigure[]{\includegraphics[width=.21\textwidth]{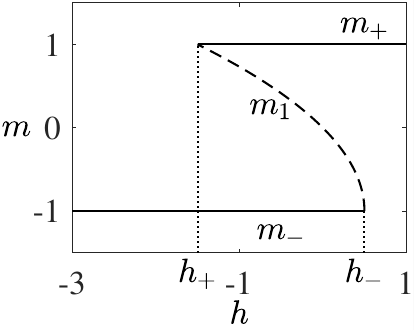}\label{fig:FIM_bif_gamma13}}
    \subfigure[]{\includegraphics[width=.21\textwidth]{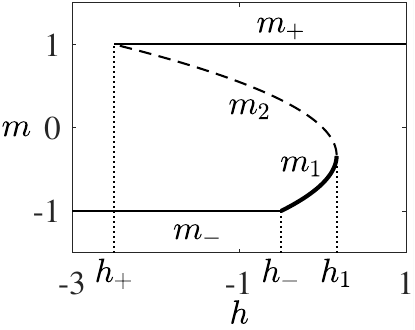}\label{fig:FIM_bif_gamma1}}
    \caption{Bifurcation diagram of equilibria on the $(h,m)$-plane in (a,b) the classical IM; and (c,d) the linear FIM. Solid branches are stable, bold branches are super-stable, and dashed branches are unstable. The feedback function $f(m)$ is (a) $1$; (b) $-1$; and (c,d) $1+\gamma m$ with (c) $\gamma=\frac{1}{3}$; and (d) $\gamma=1$. For panels (c) \& (d), the bifurcation points are $h_\pm=\mp1-\frac{3}{2}\gamma$ and $h_1=1/(6\gamma)$ for $\gamma>\frac{1}{3}$, or explicitly: (c) $h_+=-\tfrac32$ and $h_-=\tfrac12$; (d) $h_+=-\tfrac52$, $h_-=-\tfrac12$, and $h_1=\tfrac16$.}
    \label{fig:FIM_bif}
\end{figure}

In the classical IM with $f(m)=s$, where $s=1$ in the ferromagnetic case and $s=-1$ in the antiferromagnetic case, the ${\cal C}_0$ curve is the line segment $g(m)=-sm$ with endpoints $(h_\pm,m_\pm)=(\mp s,\pm1)$. In either case, there is a single MP branch $m_1$. Since $g'(m_-)=-s$, this branch is stable when $s=-1$ and unstable when $s=1$. Thus, stable MPs only exist in the antiferromagnetic case. As shown in Figs.~\ref{fig:IM_bif} \& \ref{fig:IM_bif_anti}, bistability only exists in the ferromagnetic case between $m_\pm$ when $h\in(-1,1)$.

In the linear FIM with $f(m)=1+\gamma m$, the ${\cal C}_0$ curve is the parabolic segment $g(m)=-m\left(\frac{3}{2}\gamma m+1\right)$ with endpoints $(h_\pm,m_\pm)=(\mp1-\frac{3}{2}\gamma,\pm1)$. For $\gamma\in(0,\frac{1}{3}]$, $g(m)$ is monotonic between the endpoints as shown in Fig.~\ref{fig:FIM_bif_gamma13}. In this case, there is a single unstable MP branch $m_1$. However, for $\gamma>\frac{1}{3}$, $g(m)$ has its vertex $(h_1,m)=(1/(6\gamma),-1/(3\gamma))$ between the endpoints as shown in Fig.~\ref{fig:FIM_bif_gamma1}. In this case, there is a stable MP branch $m_1$ below the vertex and an unstable MP branch $m_2$ above the vertex. Remarkably, the linear FIM can exhibit stable MPs with ferromagnetic coupling.

Hereafter, we assume $\gamma>\frac{1}{3}$ in the linear FIM such that stable MPs can exist. The existence interval for stable MP in terms of $f(m)$ is $f(m)\in\left(1-\gamma,\frac{2}{3}\right)$, whose upper bound is independent of $\gamma$, so stable MPs in the linear FIM always persist until the coupling has increased to $\frac{2}{3}$.


{\it Phase transitions.}---FIMs can exhibit four types of phase transitions on the bifurcation diagram. Type 1 is a smooth fold $h_i$, $i=1,2,\cdots$, on the ${\cal C}_0$ curve. Type 2 is a sharp fold at $h_\pm$ when $g'(m_\pm)<0$. Both are first-order phase transitions, i.e., discontinuous phase transitions or tipping points. These transitions are irreversible, i.e., they happen when $h$ varies in one direction but not the other. Both type 3 and type 4 involve a transcritical bifurcation at $h_\pm$ when $g'(m_\pm)>0$, which is a second-order phase transition, i.e., a continuous phase transition with a discontinuous first derivative. This transition is reversible, with type 3 turning $m_\pm$ into stable MP and type 4 turning stable MP into $m_\pm$.

\begin{figure}
    \centering
    \subfigure[]{\includegraphics[width=.22\textwidth]{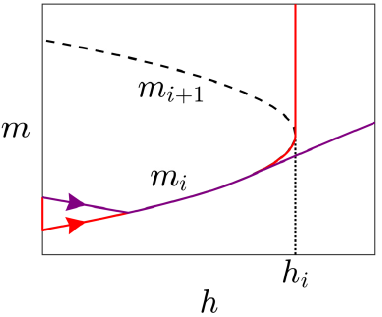}\label{fig:FIM_type1}}
    \subfigure[]{\includegraphics[width=.23\textwidth]{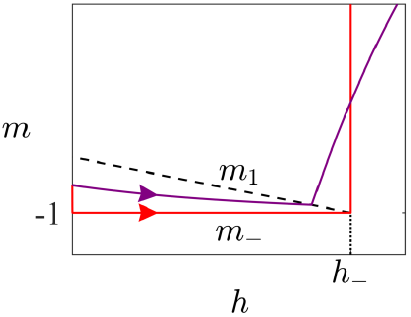}\label{fig:IM_type2}}
    \subfigure[]{\includegraphics[width=.23\textwidth]{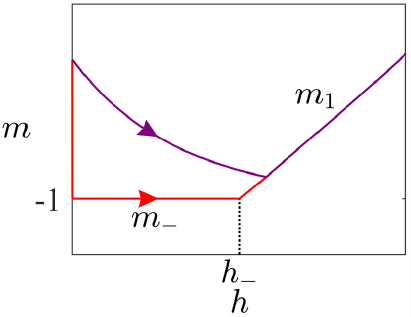}\label{fig:FIM_type3}}
    \subfigure[]{\includegraphics[width=.23\textwidth]{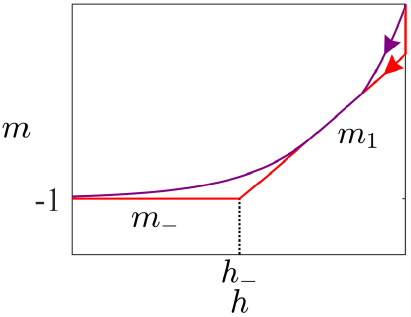}\label{fig:FIM_type4}}
    \caption{Four types of phase transitions on the $(h,m)$-plane: (a) type-1; (b) type-2; (c) type-3; and (d) type-4. For each panel, the red trajectory shows an exact transition with $h'(t)\to0$, while the magenta trajectory shows an inexact transition with $|h'(t)|$ small.}
    \label{fig:FIM_pt_type}
\end{figure}

The initial condition may not coincide with the equilibrium, and $h'(t)$ may not be infinitesimal. In the static $h'(t)\to 0$ limit, we call the transitions exact since the system follows the bifurcation diagram exactly after an initial transient. In the dynamic case with small $|h'(t)|$, we call the transitions inexact. As shown in Fig.~\ref{fig:FIM_pt_type} and associated discussions in End Matter, inexact type-1 and type-4 transitions are third-order, while inexact type-2 and type-3 transitions are second-order.

As shown in Figs.~\ref{fig:IM_bif} \& \ref{fig:IM_bif_anti}, the classical IM can only exhibit type-2 transitions at $h_\pm$ in the ferromagnetic case and type-3 and type-4 transitions at $h_\pm$ in the antiferromagnetic case. The linear FIM can exhibit all four types of transitions, including type-1 at $h_1$, type-2 at $h_+$, and type-3 and type-4 at $h_-$; see Fig.~\ref{fig:FIM_bif_gamma1}. Thus, the linear FIM provides a minimal model for type-1, type-3, and type-4 transitions with fully ferromagnetic coupling.

In applications, it is crucial to identify the sequence of phase transitions as a control parameter varies. The signature of the linear FIM as $h$ increases is the three-stage phase transition $m_-\xrightarrow{3}m_1\xrightarrow{1}m_+$, where the transition type is shown above each arrow. In practice, one may need not only to generate data from a model but also to develop a model from data. In real-world data, a system may exhibit a lower phase $ m_{-} = - 1$ for $h$ below a critical value, an upper phase $ m_{+} = + 1$ for $h$ above a certain value, and intermediate phase(s) on a compact interval(s) of $h$, all of which are stable. If each $|m|<1$ maps to at most one $h$ value in the intermediate phase(s), one can interpolate the known data into a bifurcation diagram $h=g(m)$ for all $|m|<1$ that describes phase transformation from $m_-$ to $m_+$ via stable and unstable intermediate phase(s). The challenge is to construct a phenomenological model of the system relying only on this bifurcation diagram.

We argue that FIMs can be the method of choice for this challenge. To construct a feedback function $f$ without singularities, one can first shift $h$ if needed such that $g(0)=0$ and then use Eqs.~(\ref{eq:hat-H}) \& (\ref{eq:hat-H-prime}) with $h=0$ to get
\begin{equation}\label{eq:g-to-f}
    f(m)=-\frac{2}{m^2}\int_0^mg(u)du.
\end{equation}
Note that although mean-field FIMs can model arbitrary bifurcation diagrams, finite-dimensional FIMs are needed to model spatial patterns.

Despite the generality of mean-field FIMs, we expect them to be most useful when the feedback function $f$ and the bifurcation diagram $g$ are both simple. Thus, the bifurcation diagram in Fig.~\ref{fig:FIM_bif_gamma1} predicted by the linear FIM should find most applications.

In the general FIM, one can derive analytically the Maxwell point $h_*$ between the two stable branches $m_L$ and $m_U$, i.e., the $h$ value such that $m_L$ and $m_U$ have equal Hamiltonian. In the linear FIM, one can then show that stable MPs can be ground states only when $\gamma>1$, i.e., only when the coupling is partly antiferromagnetic; see Supplemental Material~\cite{SM}.

\begin{figure}
    \centering
    \includegraphics[width=0.5\linewidth]{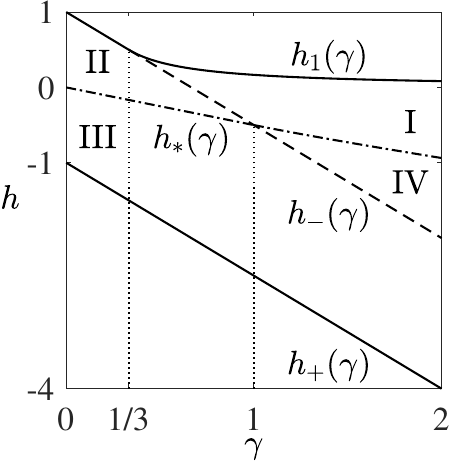}
\caption{Phase diagram of the linear FIM on the $(\gamma,h)$-plane at zero temperature. The bistable regime has a lower boundary $h_+(\gamma)$ and a two-piece upper boundary consisting of $h_-(\gamma)$ for $\gamma\in[0,\tfrac13]$ and $h_1(\gamma)$ for $\gamma>\tfrac13$. The locus of second-order phase transitions is $h_-(\gamma)$ for $\gamma>\tfrac13$. The Maxwell curve $h_*(\gamma)$ is linear for $\gamma\in[0,1]$ and nonlinear for $\gamma>1$; see Supplemental Material~\cite{SM}. These two curves $h_-(\gamma)$ and $h_*(\gamma)$ divide the bistable regime into four regions denoted as regions I--IV.}
    \label{fig:FIM_gh_T0}
\end{figure}

The phase diagram of the linear FIM at zero temperature is shown in Fig.~\ref{fig:FIM_gh_T0}. The locus of second-order phase transitions $h_-(\gamma)$ and the Maxwell curve $h_*(\gamma)$ divide the bistable regime into four regions. Below $h_-(\gamma)$, i.e., in regions II and III, $m_-$ and $m_+$ are bistable. Above $h_-(\gamma)$, i.e., in regions I and IV, $m_1$ and $m_+$ are bistable. Above $h_*(\gamma)$, i.e., in regions I and II, $m_+$ is the ground state. Below $h_*(\gamma)$, $m_-$ is the ground state in region III and $m_1$ is the ground state in region IV.

In Ref.~\cite{ma2025dynamics}, we study the dynamics of the linear FIM in systems with large but finite $N$ and at any temperature. For positive temperature, the locus of second-order phase transitions disappears, and the Maxwell curve $h_*(\gamma)$ becomes smooth. Counterintuitively, temperature-induced bistability can emerge at higher temperatures, where an increase in temperature may favor bistability between two phases. 

Multistability is among the most interesting phenomena in dynamical systems \cite{pisarchik2014control}. The FIM provides a minimal model for multistable systems based solely on observable data. Such data-driven modeling of multistable systems can yield effective control strategies even with limited information. 

In finite dimensions, the FIM may exhibit spatial patterns with intricate bifurcation structures as in multistable reaction-diffusion equations \cite{knobloch2015spatial}. In the End Matter, we briefly define the finite-dimensional FIMs and reveal how their Glauber dynamics can deviate from those of the classical IM due to a self-modulation effect. A detailed analysis of the 2D FIM will be presented elsewhere \cite{ma2026dynamics}.

The statistical mechanics and dynamics of both classical \cite{Ruffo09} and quantum \cite{defenu2023long} systems with long-range (LR) interactions are active research areas  \cite{Dyson69,sak1973recursion,angelini2014relations,Behan_2017}. One can relate finite-dimensional FIMs to the long-range Ising model (LRIM). In the $d$-dimensional LRIM, the two-spin interaction exhibits a power-law decay $J_2 \propto R^{-d-s}$ where $R$ is the geodesic distance and $s>0$. Thus, the LRIM approaches the nearest-neighbor IM for large $s$ and the mean-field IM for small $s$. The interaction $J_2$ is LR for $s<2$ and SR for $s>2$, but the critical exponents are shown to cross over from LR to SR at some $s_*<2$ \cite{sak1973recursion}. One can include feedback in the LRIM by requiring that the LR limit of the feedback LRIM is the mean-field FIM. Thus, the feedback LRIM contains not only the interaction $J_2$ of the classical LRIM but also $p$-spin interactions, $p\geq3$, of the form $J_p\propto D_p^{-a_p}$ where $D_p$ is a $p$-spin distance and $a_p>0$. In the End Matter, we identify finite-dimensional FIMs with feedback LRIMs for a particular choice of $D_p$. Generally, in the feedback LRIMs, new types of crossover from LR to SR are possible depending on the choice of $D_p$. Characterizing the locus of $s$ and $a_p$'s for such crossovers is challenging. Equilibrium and dynamical behaviors near such crossovers are rich subjects for further exploration.

The FIM elevates spin-spin coupling to an arbitrary function of magnetization, provides a platform for feedback-controlled order, and admits finite-dimensional extensions that may uncover universality classes beyond the classical IM. Because similar micro-macro feedback loops govern phase transformations in socio-economic \cite{macy2024ising} and geophysical \cite{banwell2023physics} systems, the FIM offers a unifying framework to mimic phenomena such as market crashes, volatility clustering, and climate tipping points. With its combination of analytical tractability and real-world relevance, the FIM is poised to become a benchmark model for studying feedback-driven phase transformations across disciplines.

{\it Acknowledgments.} This research was made possible by a Research-in-Groups programme funded by the International Centre for Mathematical Sciences, Edinburgh. I.S. gratefully acknowledges support by the NSF through Grant No. PHY-2102906.

\bibliography{references}

\newpage
\begin{center}
    {\bf End Matter}
\end{center}

{\it Possible bifurcation diagrams of equilibria.}---A bifurcation diagram of equilibria in the general FIM is defined by the curve $h=g(m)$, $m\in[-1,1]$, where $g$ can be any continuous function. By Weierstrass approximation theorem, $g$ can be well approximated by a polynomial. This polynomial approximation is practically useful since it yields a polynomial feedback function $f$ via Eq.~(\ref{eq:g-to-f}). This provides a systematic approach to decompose any bifurcation diagram into multi-spin interactions.

In Fig.~\ref{fig:FIM_illu}, the functions $g$ used in each panel are: (a) $g(m)=\sin(-\tfrac76\pi m)$; (b) $g(m)=\sin(\tfrac76\pi m)$. Thus, the bifurcation points are: (a) $h_1=-1$, $h_+=-\tfrac12$, $h_-=\tfrac12$, and $h_2=1$; (b) $h_2=-1$, $h_-=-\tfrac12$, $h_+=\tfrac12$, and $h_1=1$.

{\it Dynamical phase transitions.}---Figure \ref{fig:FIM_pt_type} shows the four types of phase transitions in a time-varying magnetic field. For each type, we show an exact transition with $h'(t)\to0$ and an inexact transition with $|h'(t)|$ small.

Although exact type-1 and type-2 transitions are both first-order, their inexact versions have different orders. An inexact type-1 transition is a separation from a stable MP branch predicted by either the lower bound or the upper bound of Eq.~(\ref{eq:trap-cond}). This is a third-order phase transition, i.e., a continuous phase transition with a discontinuous second derivative; see Fig.~\ref{fig:FIM_type1}. An inexact type-2 transition is a deflection off an unstable MP branch, which is a second-order phase transition; see Fig.~\ref{fig:IM_type2}.

Although exact type-3 and type-4 transitions are both second-order, their inexact versions have different orders. An inexact type-3 transition is a merger onto a stable MP branch, which is second-order; see Fig.~\ref{fig:FIM_type3}. An inexact type-4 transition involves both the trajectory $(h(t),m(t))$ that is locally exponential and the stable MP branch that is locally linear. If these two curves do not intersect, then there are no transitions. Otherwise, there is a separation from the stable MP branch, which is third-order; see Fig.~\ref{fig:FIM_type4}.

In Fig.~\ref{fig:FIM_pt_type}, the governing equation (\ref{eq:dmdt-general}) is solved using the MATLAB solver \texttt{ode15s} for stiff ordinary differential equations at an extremely low temperature $\beta^{-1} = 10^{-6}$. The magnetic field is $h(t) = h_L+(h_U-h_L)\Omega t$ and the initial condition is $m(t=0)=m_0$, where $h_L$, $h_U$, $\Omega$, and $m_0$ are constants. The parameters used in each panel are: (a) $\gamma = 1$, $h_L = 0.14$, $h_U = 0.175$, $m_0 = -0.42$, and $\Omega = 10^{-4}$ (red) and $\Omega = 10^{5/6}$ (magenta); (b) $\gamma=0$, $h_L = 0.8$, $h_U = 1.04$, $m_0 = -0.9$, and $\Omega = 10^{-4}$ (red) and $\Omega = 10^{-1/4}$ (magenta); (c) $\gamma = 1$, $h_L = -0.6$, $h_U = -0.4$, $m_0 = -0.95$, and $\Omega = 10^{-4}$ (red) and $\Omega = 10^{-1/2}$ (magenta); (d) $\gamma = 1$, $h_L = -0.6$, $h_U = -0.4$, $m_0 = -0.93$, and $\Omega = 10^{-4}$ (red) and $\Omega = 10^{-2/3}$ (magenta).

{\it Finite-dimensional FIM.}---For a finite-dimensional FIM, the coordination number of each spin is proportional to the dimensionality. We let the feedback coupling for a bond $\langle i,j\rangle$ depend on the average magnetization $m_{\langle i,j\rangle}$ in its $r$-neighborhood ${\cal N}_{\langle i,j\rangle,r}$. We define the $r$-neighborhood as the union of the two balls of radius $r$ and centers $i$ and $j$, but excluding these two centers, 
\begin{equation}
    {\cal N}_{\langle i,j\rangle,r}=\{k:\min(d_{i,k},d_{j,k})\leq r\}\backslash\{i,j\},
\end{equation}
where $d_{i,j}$ denotes the geodesic distance between sites $i$ and $j$. Thus, the average magnetization is
\begin{equation}
\label{eq:mi_def}
    m_{\langle i,j\rangle}=\frac{1}{\#({\cal N}_{\langle i,j\rangle,r})}\sum_{k\in{\cal N}_{\langle i,j\rangle,r}}s_k,
\end{equation}
where $\#(\cdot)$ counts the number of elements. We define the Hamiltonian as the following generalization of Eq.~(\ref{ham-F}):
\begin{equation}
\label{eq:H_FIM_bond}
H_\text{FIM}=-h\sum_is_i-\frac{1}{N}\sum_{\langle i,j\rangle}f(m_{\langle i,j\rangle})s_is_j.
\end{equation}
We call the FIM local if $r=1$, regional if $2\leq r<\infty$, and global if $r\to\infty$.

The simplest such FIM with $f$ linear contains three-spin interactions of the form $\{\langle i,j\rangle,k\}$ where $\langle i,j\rangle$ is a bond and $k$ is in the $r$-neighborhood of this bond. Generally, such FIM contains $p$-spin interactions, $3\leq p\leq\#({\cal N}_{\langle i,j\rangle,r})+2$, of the form $\{\langle i,j\rangle,k_1,\cdots,k_{p-2}\}$ where $\langle i,j\rangle$ is a bond and $k_1,\cdots,k_{p-2}$ are in the $r$-neighborhood of this bond.

It is also useful to rewrite the second term of the Hamiltonian (\ref{eq:H_FIM_bond}) as a summation over all $p$-spin combinations ${\cal S}_p\equiv\{k_1,\cdots,k_p\}$ instead of all bonds. For each ${\cal S}_p$, the $p$-spin interaction is proportional to the number of bonds in ${\cal S}_p$ such that all other spins in ${\cal S}_p$ are in the $r$-neighborhood of this bond, i.e.,
\begin{equation*}
    B({\cal S}_p)=\#(\{\{i,j\}\subset{\cal S}_p:d_{i,j}=1,{\cal S}_p\backslash\{i,j\}\subseteq{\cal N}_{\langle i,j\rangle,r}\}).
\end{equation*}
We call $B$ the bond-counting function. As $r\to\infty$, $B({\cal S}_p)$ simply counts the number of bonds in ${\cal S}_p$. In graph theory, this number is often called the internal edge count or the edge count of the induced subgraph. Recalling that $p$-spin interactions in the feedback LRIM take the form $J_p\propto D_p^{-a_p}$, we can identify the finite-dimensional FIM with the feedback LRIM for the particular choice $D_p=1/B({\cal S}_p)$ and $a_p=1$.

Now we compare the Glauber dynamics of FIM and the classical IM with Hamiltonian $H_\text{IM}=-\G{\cal M}-J{\cal I}$, 
where ${\cal M}=\sum_is_i$ and ${\cal I}=\frac{1}{N}\sum_{\langle i,j\rangle}s_is_j$. The change of the Hamiltonian due to flipping spin $s_k$ is
\begin{equation}
\label{eq:Delta_H_IM}
    \Delta(H_\text{IM})_k=-\G\Delta{\cal M}_k-J\Delta{\cal I}_k,
\end{equation}
where $\Delta{\cal M}_k=-2s_k$ and $\Delta{\cal I}_k=(-2s_k/N)\sum_{i:\langle i,k\rangle}s_i$.

For the FIM, the change of the Hamiltonian defined by \eqref{eq:H_FIM_bond} due to flipping spin $s_k$ is
\begin{equation}
\label{eq:Delta_H_FIM}
\begin{aligned}
    &\Delta(H_\text{FIM})_k=-h\Delta{\cal M}_k+\frac{2s_k}{N}\sum_{i:\langle i,k\rangle}f(m_{\langle i,k\rangle})s_i\\
    &-\Delta{\cal M}_k\cdot\frac{1}{N}\sum_{\langle i,j\rangle:k\in{\cal N}_{\langle i,j\rangle,r}}\frac{f'(m_{\langle i,j\rangle})}{\#({\cal N}_{\langle i,j\rangle,r})}s_is_j.
\end{aligned}
\end{equation}
The second term of Eq.~(\ref{eq:Delta_H_FIM}) is $-\Delta{\cal I}_k$ but with the sum over nearest neighbors $i$ weighted by $f(m_{\langle i,k\rangle})$. The third term of Eq.~(\ref{eq:Delta_H_FIM}) is $-\Delta{\cal M}_k{\cal I}$ but with the sum restricted to those bonds $\langle i,j\rangle$ whose $r$-neighborhood includes $k$ and weighted by $f'(m_{\langle i,j\rangle})/\#({\cal N}_{\langle i,j\rangle,r})$. For FIM with small $r$, e.g., the local FIM, the arguments of both $f$ and $f'$ vary rapidly in $k$, so the Glauber dynamics deviate from those of the classical IM.

However, for FIM with $r\to\infty$, i.e., the global FIM, we have $\#({\cal N}_{\langle i,j\rangle,r})=N$ and $m_{\langle\cdot,\cdot\rangle}=m$ in Eq.~(\ref{eq:Delta_H_FIM}), where $m$ is the global magnetization. In this case, $\Delta(H_\text{FIM})_k$ becomes $\Delta(H_\text{IM})_k$ with the effective parameters
\begin{equation}
\label{eq:HJ_global}
    \G(m)=h+\frac{f'(m)}{N}{\cal I},\quad J(m)=f(m).
\end{equation}
Thus, metastable states of the global FIM, i.e., those with $\Delta(H_\text{FIM})_k\geq0$ for all $k$, reduce to metastable states of the effective IM. Generally, the Glauber dynamics of the global FIM reduce to those of the effective IM with state-dependent parameters in Eq.~(\ref{eq:HJ_global}).

Like the global FIM, the mean-field FIM can be identified with the effective IM. Indeed, since ${\cal I}=\tfrac12Nm^2$ and $\Delta{\cal I}_k=-2s_km$ in the mean-field case, $\Delta(H_\text{IM})_k$ agrees with ${\cal H}'(m)$ in Eq.~(\ref{eq:hat-H-prime}). Thus, MPs of the mean-field FIM, i.e., solutions to ${\cal H}'(m)=0$, agree with MPs of the effective IM. However, the stability of MPs depends on the second derivatives and can differ between these two models; e.g., stable MPs in Fig.~\ref{fig:FIM_bif_gamma1} are unstable in the effective IM.

Finally, for regional FIM with large $r$, $m_{\langle i,k\rangle}$ can be approximated by the average magnetization $m_k$ in a ball of radius $r$ and center $k$. Thus, $\Delta(H_\text{FIM})_k$ can be approximated by $\Delta(H_\text{IM})_k$ with the effective magnetic field
\begin{equation}
    \G_k=h+\frac{1}{N}\sum_{\langle i,j\rangle:k\in{\cal N}_{\langle i,j\rangle,r}}\frac{f'(m_{\langle i,j\rangle})}{\#({\cal N}_{\langle i,j\rangle,r})}s_is_j,
\end{equation}
and the effective coupling $J_k=f(m_k)$. Moreover, both $\G_k$ and $J_k$ vary slowly in $k$ on the $O(r)$ length scale. Thus, the microscopic Glauber dynamics of FIM below the $O(r)$ length scale reduce to those of the classical IM with state-dependent parameters. However, the macroscopic Glauber dynamics of FIM exhibit a self-modulation effect on the $O(r)$ length scale due to the slow variation of the effective parameters $\G_k$ and $J_k$. This may enable domain walls between metastable states in the classical IM at different parameters, e.g., between a ferromagnetic pattern and an antiferromagnetic one.

\end{document}